\pdfoutput=1

\documentclass[11pt]{article}

\PassOptionsToPackage{hyperfootnotes=false}{hyperref}
\usepackage[final]{acl}

\usepackage{times}
\usepackage{latexsym}

\usepackage{booktabs}
\usepackage{tabularx}
\usepackage{multirow}
\usepackage{xcolor}
\usepackage{colortbl}
\usepackage{array}
\usepackage{amsmath}
\usepackage{amsthm}
\usepackage{makecell}
\usepackage{subcaption}

\newtheorem{definition}{Definition}

\newcolumntype{C}{>{\centering\arraybackslash}p{0.07\textwidth}}

\usepackage[T1]{fontenc}
\usepackage[utf8]{inputenc}
\usepackage{microtype}
\usepackage{inconsolata}
\usepackage{graphicx}
\usepackage{pifont}

\newcommand{\equalcontribsymbol}{\ding{96}}
\newcommand{\correspondingsymbol}{\ding{72}}

\title{Position: Recommender Systems Should Move Beyond\\Platform-Centric Ranking toward Personal Agent-Mediated Recommendation}

\author{
  Haohan Yuan\textsuperscript{1}\thanks{Equal contribution.} \quad
  Peng He\textsuperscript{2,3}\footnotemark[1]\thanks{Corresponding author.} \quad
  Dan Zhang\textsuperscript{4} \quad
  Jianpeng Liang\textsuperscript{5} \quad
  Junning Zhu\textsuperscript{6} \\
  \textsuperscript{1}University of North Carolina at Charlotte \\
  \textsuperscript{2}Tsinghua University \quad
  \textsuperscript{3}Sophon.ai \quad
  \textsuperscript{4}Beijing University of Posts and Telecommunications \\
  \textsuperscript{5}University of California San Diego \quad
  \textsuperscript{6}Beijing Normal-Hong Kong Baptist University \\
  \texttt{hyuan3@charlotte.edu} \quad
  \texttt{hepeng@tsinghua-wx.org} \quad
  \texttt{2025110629@bupt.cn} \\
  \texttt{jil652@ucsd.edu} \quad
  \texttt{t330025113@mail.bnbu.edu.cn}
}

\begin{document}
\renewcommand{\thefootnote}{\ifcase\value{footnote}\or\equalcontribsymbol\or\correspondingsymbol\else\arabic{footnote}\fi}
\maketitle
\renewcommand{\thefootnote}{\arabic{footnote}}

\begin{abstract}
Recommender systems are usually framed as ranking systems: platforms observe
users, construct candidate sets, and select items on their behalf.
This framing hides a deeper allocation of control, in which platforms also
determine candidate access, evidence boundaries, explanations, and the path
from user need to recommended output.
We argue that the next bottleneck in recommendation is not only preference
modeling, but control over evidence acquisition and disclosure.
We argue for \textbf{Personal Agent-Mediated Recommendation} (PAMR), a paradigm
in which a user-facing personal agent represents the user in discovering,
filtering, aggregating, and governing recommendation evidence across
distributed sources.
The central shift is not simply from one ranking model to another, but from
platform-side item ranking to user-side evidence mediation.
As a position paper, we define PAMR as a new recommendation paradigm, establish
its boundary criteria, identify its core mediation decisions, and propose a
mediation-centered evaluation framework.
A proof-of-concept study on hard Yelp restaurant recommendation tasks further
shows that, under a shared LLM ranker, source selection and controlled
disclosure provide the strongest observed
utility--traceability--exposure--cost operating point.
\end{abstract}

\section{Introduction}

\begin{figure}[!t]
\centering
\includegraphics[width=0.90\columnwidth]{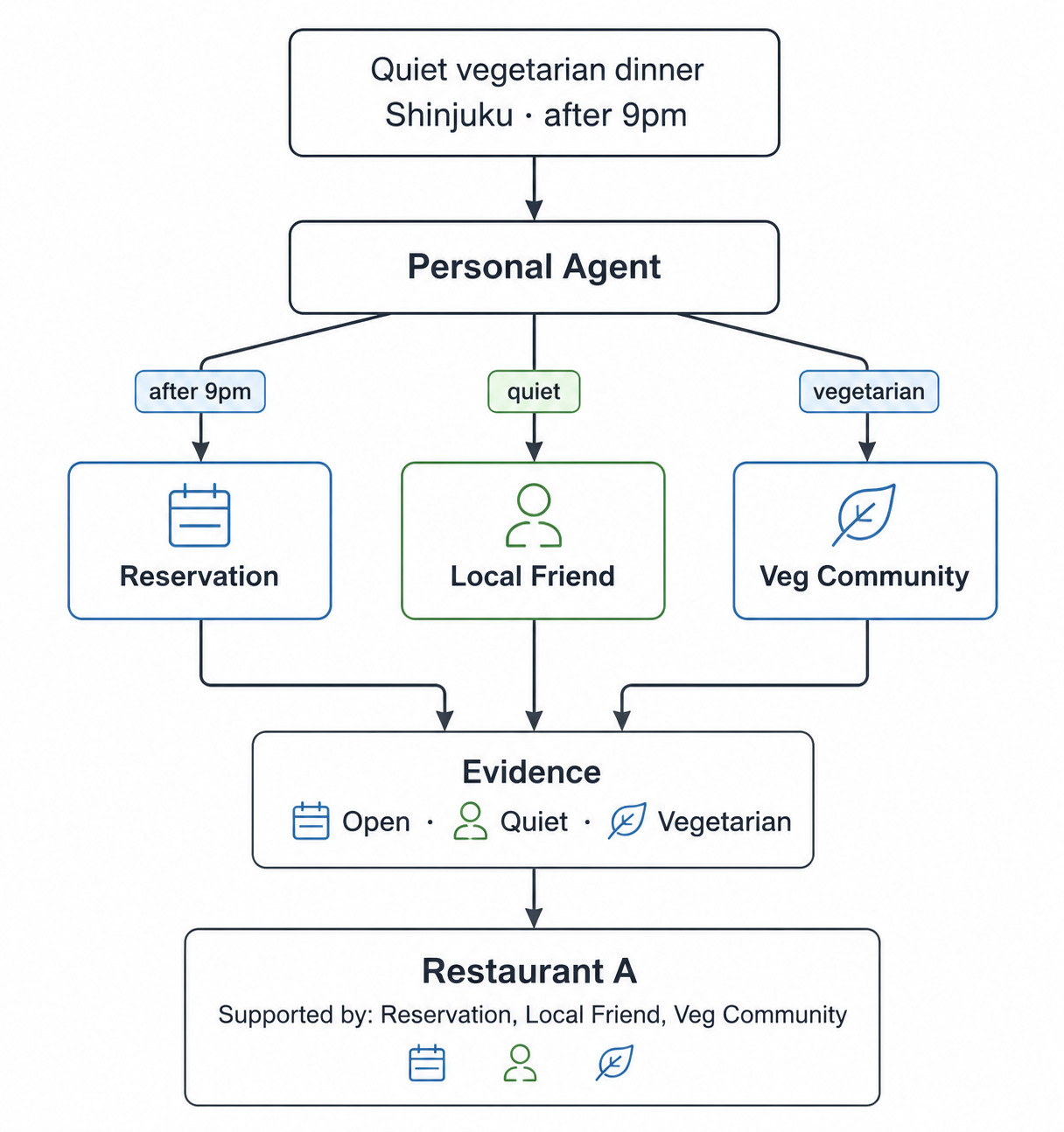}
\caption{Selective evidence mediation for a restaurant request. The personal
agent routes source-relevant constraints and combines provenance-linked
evidence into a recommendation.}
\label{fig:mediation_shift}
\vspace{-0.8em}
\end{figure}

Recommender systems are usually framed as ranking systems: a platform observes
user behavior, models preferences, constructs a candidate set, and selects
items on the user's behalf \citep{adomavicius2005toward, ricci2015recommender}.
This framing has supported decades of progress, from matrix factorization and
neural recommenders to social and privacy-preserving recommendation
\citep{koren2009matrix, he2017neural, ma2008sorec, fan2019graph,
ammad2019federated}.
Yet these technically different systems often preserve the same allocation of
control: the platform bounds the candidate space, decides which evidence
matters, controls explanation, and mediates the path from user need to ranked
output.
This control is often invisible. Users see ranked outputs, but not the
evidence boundaries, source-selection decisions, disclosure choices, or
mediation policies that produced them.
The next bottleneck in recommendation is therefore not only preference
modeling, but control over evidence acquisition and disclosure.
Figure~\ref{fig:mediation_shift} illustrates the shift from platform control
over candidates and ranking to personal-agent mediation over distributed
evidence.

Consider a user asking for a quiet, vegetarian-friendly dinner near Shinjuku
after 9pm.
No single ranking interface necessarily contains all relevant evidence:
opening hours, reservation availability, local atmosphere, dietary fit, and
social trust may live across different services and people.
A personal agent might ask a reservation platform only about availability, a
local friend only about atmosphere, and a vegetarian community only about
dietary fit, while recording which context was disclosed to each source.
It might then aggregate the returned evidence, preserve disagreement, and let
the user say that future dinner requests should prioritize local friends over
ad-heavy services.
The recommendation problem is therefore not only which restaurant should be
ranked first, but who should be asked, what context should be shared, which
sources should be trusted, how disagreement should be preserved, and how much
control the user should retain.

We call this shift \emph{user-side evidence mediation}.
It changes four assumptions of platform-centric recommendation:
\emph{evidence boundary} moves beyond platform inventory;
\emph{decision locus} moves to a user-inspectable mediator;
\emph{traceability} moves from opaque ranked lists to provenance, disclosure
traces, uncertainty, and disagreement; and \emph{adaptation} moves from item
preference learning to mediation-policy learning.

Personal language agents make this reframing timely, while also making the
risk sharper.
Large language models now power agents that plan, use tools, communicate, and
act on a user's behalf \citep{wang2024survey, park2023generative}, while a
growing body of work applies LLMs directly to recommendation
\citep{wu2024survey, hou2024large, huang2023recommender, zhang2024generative}.
Increasingly, such agents are imagined as \emph{personal}: persistent,
user-facing systems that hold user context and interface with information,
services, and other agents \citep{li2024personal}.
But without explicit user-side evidence mediation, personal-agent recommenders
can reproduce platform-centric control through conversational interfaces: the
system may sound user-controlled while the candidate boundary, evidence
sources, disclosure choices, and adaptation policy remain controlled
elsewhere.
From the perspective of PAMR, LLM recommenders that only improve ranking or
explanation are solving the wrong problem if they leave evidence boundaries
platform-controlled.
Existing work provides important fragments: LLM and agentic recommenders
often improve reasoning or ranking
\citep{huang2023recommender, zhang2024generative, wang2024recmind}; social,
federated, and conversational recommendation address social signals, locality,
or interaction; and general multi-agent systems study collaboration as a broad
mechanism \citep{wu2023autogen, guo2024large}.
What remains missing is a recommendation-specific research object in which
mediation control itself is explicit, traceable, and adaptive.

\paragraph{Position.}
We argue that recommender systems should move beyond platform-centric ranking
toward
\textbf{Personal Agent-Mediated Recommendation} (PAMR): a paradigm in which a
user-facing personal agent mediates distributed recommendation evidence under
user-inspectable policies for utility, cost, latency, privacy, trust, and
control.
Multi-agent communication is the mechanism, but the research contribution is
the new allocation of mediation.
A PAMR system should therefore be evaluated not only by whether it recommends
better items, but also by whether it helps users understand and control where
evidence comes from, what is disclosed, whom to trust, how disagreement is
handled, and how future mediation changes.

\begin{figure*}[t]
\centering
\includegraphics[width=0.90\textwidth]{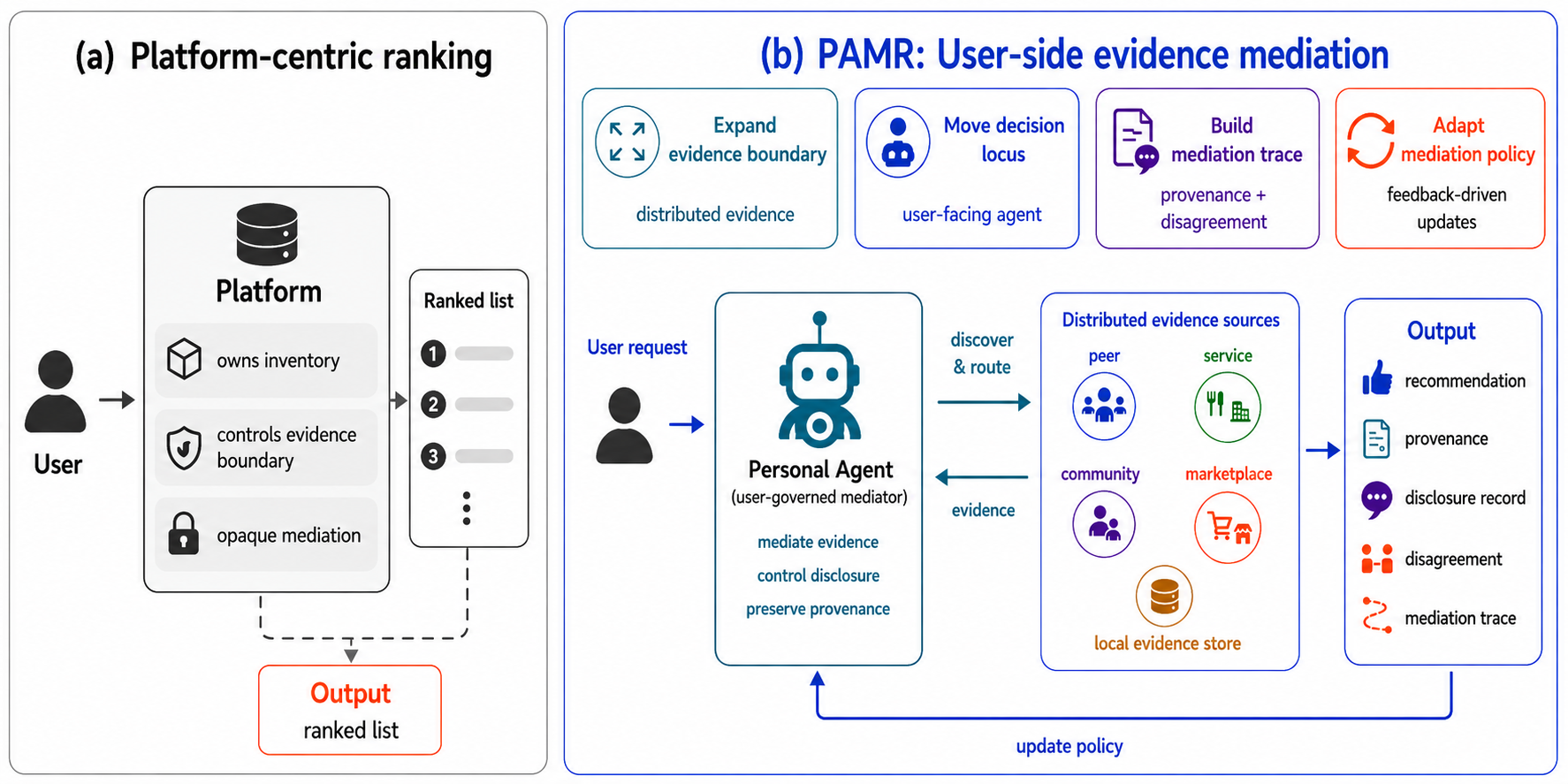}
\caption{Conceptual overview of the shift beyond platform-centric ranking
toward
Personal Agent-Mediated Recommendation (PAMR). Platform-centric systems own
the inventory, control evidence boundaries, and return ranked lists, whereas
PAMR centers a user-facing personal agent that discovers distributed evidence,
preserves provenance and disagreement, and adapts future mediation policy.}
\label{fig:pamr_conception}
\end{figure*}

We develop this position by treating mediation control as a first-class
object of recommender-systems research, rather than as an implementation
detail hidden behind ranking or conversational interfaces.
Our contributions are:

\begin{itemize}
  \item We define PAMR through four shifts: evidence boundary, decision
  locus, traceability, and mediation-policy adaptation, and use boundary
  criteria to distinguish it from adjacent agentic recommendation paradigms.

  \item We formulate PAMR's mediation policy as a set of core decisions:
  source discovery and routing, privacy-budgeted disclosure,
  provenance-preserving aggregation, and feedback-driven adaptation.

  \item We provide a controlled proof-of-concept study showing that source
  selection and disclosure control make complementary contributions under a
  shared LLM ranking layer.
\end{itemize}

\section{Related Paradigms and Gaps in User-Side Mediation}
\label{sec:background}

PAMR builds on recommender systems, conversational systems, personal agents,
and multi-agent coordination, but it is not simply their combination.
The key question is whether a paradigm changes the allocation of mediation:
who controls the evidence boundary, who makes request-time mediation
decisions, whether the process is traceable, and what kind of behavior adapts
over time.
Recent recommender agents have made rapid progress on reasoning, memory,
tool use, conversation, and user simulation.
PAMR draws from these advances, but treats user-side evidence mediation as
the shared object that connects source access, disclosure, provenance,
disagreement, and policy adaptation.

\paragraph{Platform-bounded recommendation.}
Collaborative filtering, content-based recommendation, matrix factorization,
and neural recommenders model users and items to produce rankings
\citep{koren2009matrix, he2017neural}.
Social recommendation broadens the available signals through relationships
\citep{ma2008sorec, fan2019graph}, while federated recommendation brings
privacy and locality into model training
\citep{ammad2019federated, yang2020federated}.
These paradigms remain foundational, but they usually assume that candidates
and evidence are already available within a platform-controlled inventory.
They broaden signals or training arrangements without necessarily changing
request-time control over evidence acquisition: which people, agents,
services, communities, or platforms should be queried, what should be
disclosed to each, and how source usefulness should be learned.
Standard ranking evaluation can measure which item is preferred within a
given candidate set, but not whether the user could inspect or alter the
evidence boundary that produced it \citep{herlocker2004evaluating}.

\paragraph{Reasoning and ranking with LLMs.}
LLM-based recommendation improves how preferences, item descriptions, and
explanations are represented and ranked
\citep{wu2024survey, hou2024large, bao2023tallrec, huang2023recommender,
zhang2024generative, wang2024recmind}.
Recent reasoning-oriented systems go further: CoT-style recommendation uses
personalized reasoning traces to improve retrieval and ranking
\citep{wang2025cotrec}, R$^2$ec frames recommendation as reasoning-enhanced
item prediction \citep{you2025r2ec}, and autonomous reasoning approaches such
as RecZero optimize reasoning before recommendation \citep{kong2025think}.
These methods strengthen a crucial capability, but their primary object is
still how a model interprets or ranks evidence.
Better reasoning over a fixed or system-defined evidence boundary does not
by itself reallocate control over that boundary.
PAMR therefore treats reasoning as one component of mediation rather than as
the mediation process itself.

\paragraph{Conversational and tool-using recommendation agents.}
Conversational recommenders make interaction central
\citep{sun2018conversational, lei2020estimation, gao2021advances}, and work
on conversational information acquisition optimizes questions through
value-of-information and bandit objectives \citep{li2025evoi}.
Recent agentic systems expand this direction: ChatCRS incorporates external
knowledge retrieval and goal guidance \citep{ding2025chatcrs}, tunable
proactive recommendation agents adapt long-term proactive behavior
\citep{li2025proactive}, HARPO uses hierarchical agentic reasoning for
user-aligned conversational recommendation \citep{wang2026harpo}, and
ReasonRec combines multimodal reasoning with selective tool delegation
\citep{chen2026reasonrec}.
These systems show that interaction, external knowledge, and delegation are
becoming active recommendation decisions.
However, they usually optimize recommendation accuracy, dialogue quality,
goal progression, or agent policy, while source governance and per-recipient
context disclosure remain implicit.
PAMR asks not only what to ask or which tool to call, but which evidence
providers should receive which context, how source-specific provenance should
be preserved, and how users can revise the resulting mediation policy.

\paragraph{Personal agents, memory, and user simulation.}
The closest precursor to PAMR is iAgent, which proposes a
user-agent-platform paradigm where an LLM agent shields the user from a
platform recommender through user instructions, external tools, and personal
feedback memory \citep{xu2025iagent}.
PAMR shares this concern with user agency, but moves the unit of control from
instruction-aware reranking to evidence acquisition, per-source disclosure,
provenance-bearing aggregation, and mediation-policy adaptation.
Other recent work brings memory and simulation into recommender agents:
AgentCF++ uses memory-enhanced agents for popularity-aware cross-domain
recommendation \citep{sheng2025agentcfpp}, MemRec builds collaborative
memory for agentic recommendation \citep{liu2026memrec}, AgentRecBench
benchmarks planning, reasoning, tool use, and memory in recommender agents
\citep{shang2025agentrecbench}, and AlignUSER improves synthetic user agents
for recommender evaluation \citep{lin2026alignuser}.
These works move closer to persistent personal agents, but memory, profiles,
peer signals, and simulated users are often treated as system-managed model
context rather than user-governed evidence relationships.

\paragraph{Evidence accountability and ecosystem control.}
Another adjacent direction moves beyond unsupported ranked lists.
TRACE, for example, requires tourism recommendations to cite accountable
review evidence and support recovery after user rejection
\citep{zhao2026trace}.
Privacy-aware LLM recommendation similarly studies how to preserve utility
while limiting sensitive information exposure
\citep{khezresmaeilzadeh2025privacy}.
Ecosystem work also challenges monolithic platform recommendation:
decoupled recommender systems explore designs in which users may choose among
recommendation algorithms or middleware outside a single platform
\citep{buhayh2025decoupled}, while mechanism-design work shows that
advertisers and providers may strategically shape evidence, rankings, and
assistant interactions \citep{bergemann2026advertiser}.
These works support PAMR's emphasis on accountability, privacy, and
ecosystem control, but they usually treat these concerns as separate
components.
PAMR names their intersection as the recommendation problem itself:
deciding which evidence to seek, what context to disclose, how to preserve
provenance, how to handle disagreement, and how future mediation should
adapt.
The next section turns this argument into a definition and boundary criteria:
many systems are agentic, conversational, or multi-source, but only some make
user-side evidence mediation explicit enough to count as PAMR.

\section{Defining Personal Agent-Mediated Recommendation}
\label{sec:definition}

We now define Personal Agent-Mediated Recommendation (PAMR) as a
recommendation setting with explicit user-side evidence mediation.
The goal of this section is not only to name the paradigm, but to make it
distinguishable from LLM rankers, platform chatbots, tool-using assistants,
and general multi-agent systems.
Figure~\ref{fig:pamr_conception} summarizes the conceptual contrast between
platform-centric ranking and PAMR that the following definition makes precise.
Table~\ref{tab:pamr_boundary} situates PAMR against representative recent
recommender agents through these mediation criteria.

\begin{table*}[t]
\centering
\scriptsize
\setlength{\tabcolsep}{3pt}
\renewcommand{\arraystretch}{1.08}
\begin{tabularx}{\textwidth}{
  >{\raggedright\arraybackslash}p{0.14\textwidth}
  >{\raggedright\arraybackslash}p{0.16\textwidth}
  >{\raggedright\arraybackslash}p{0.16\textwidth}
  >{\raggedright\arraybackslash}p{0.15\textwidth}
  >{\raggedright\arraybackslash}X
  >{\raggedright\arraybackslash}p{0.15\textwidth}}
\toprule
\textbf{Work} & \textbf{Primary object} & \textbf{Evidence boundary} &
\textbf{Decision locus} & \textbf{Traceability} & \textbf{Adaptation} \\
\midrule
Personalized-information LLM Rec. (SIGIR'25) \citep{wang2025cotrec}
& Personalized reasoning and ranking
& Platform history and items
& Model/ranker
& Reasoning output
& Preference representation \\

ChatCRS (NAACL'25) \citep{ding2025chatcrs}
& Knowledge-grounded conversation
& External knowledge base
& Retrieval and goal agents
& Knowledge grounding
& Dialogue goals \\

iAgent (ACL'25) \citep{xu2025iagent}
& User instruction and reranking shield
& Platform list + external knowledge
& User-facing reranker
& Self-reflection
& Individual profile and feedback \\

AgentCF++ (SIGIR'25) \citep{sheng2025agentcfpp}
& Cross-domain user-agent memory
& Platform interaction data
& Simulated user agent
& Memory state
& Agent memory \\

R$^2$ec (NeurIPS'25) \citep{you2025r2ec}
& Reasoning-enhanced item prediction
& Fixed recommendation corpus
& Recommender model
& Reasoning chain
& Model policy \\

HARPO (ACL'26) \citep{wang2026harpo}
& User-aligned conversational decisions
& Dialogue and candidate space
& Agent policy
& Deliberative reasoning
& Recommendation policy \\

MemRec (ACL'26) \citep{liu2026memrec}
& Collaborative memory
& Community memory graph
& System memory manager
& Distilled context
& Collaborative memory \\

ReasonRec (ACL'26) \citep{chen2026reasonrec}
& Multimodal reasoning and tool delegation
& Multimodal data + specialists
& Model-controlled delegation
& Rationale and uncertainty
& Delegation policy \\

\textbf{PAMR}
& \textbf{Evidence mediation}
& \textbf{Distributed sources}
& \textbf{User-governable mediator}
& \textbf{Provenance, disclosure, disagreement}
& \textbf{Mediation policy} \\
\bottomrule
\end{tabularx}
\caption{Representative recent recommender systems through the
mediation-control lens. Cells summarize the primary capability explicitly
optimized or evaluated by each work; they do not imply that other
capabilities are technically impossible.}
\label{tab:pamr_boundary}
\end{table*}

\subsection{Definition and Core Criteria}

\begin{definition}[Personal Agent-Mediated Recommendation]
A recommender-system setting in which a user-facing personal agent acts on
behalf of a user to acquire, filter, govern, present, and adapt
recommendation evidence across distributed sources under mediation policies
the user can inspect, revise, or override.
The system is evaluated not only by final recommendation quality, but also by
how it mediates evidence access, context disclosure, provenance, disagreement,
and future mediation policy.
\end{definition}

The defining feature of PAMR is not the mere presence of multiple agents, but
the relocation of recommendation mediation from a platform-controlled ranking
process to a user-facing and user-governable process.
Thus an LLM ranker, chatbot recommender, or multi-agent task solver is not
PAMR unless user-side evidence mediation is part of what the system optimizes
or evaluates.
At minimum, this means users can inspect which sources were consulted, revise
disclosure boundaries, correct trust or routing preferences, and override or
revoke existing mediation policies.

\subsection{Four Defining Shifts}

We characterize PAMR through four defining shifts introduced in
Section~\ref{sec:background}.
First, the \emph{evidence boundary} extends beyond a single platform
inventory.
Second, the \emph{decision locus} moves to a user-facing mediator whose
policies can be inspected, revised, or overridden.
Third, \emph{traceability} covers sources, disclosed context, returned
evidence, disagreement, uncertainty, and aggregation rationale.
Fourth, \emph{adaptation} updates mediation policy, not only item preference.
Together, these shifts distinguish PAMR from systems that merely add tools,
chat, or LLM reranking to a platform-centric recommender.

\subsection{Boundary Cases: Components and Coupled Mediation}

Many current systems are agentic, tool-using, conversational, or multi-source,
but that does not automatically make them PAMR.
Table~\ref{tab:pamr_boundary} summarizes the boundary.
The table is not meant to rank systems or classify them permanently; rather,
it shows which mediation properties are explicit in representative recent
work.

Table~\ref{tab:pamr_boundary} reveals that recent agentic recommenders have
already introduced many ingredients required by PAMR.
Reasoning-oriented systems improve how candidate evidence is interpreted;
conversational agents retrieve external knowledge and optimize interaction
policies; personal-agent systems introduce user instructions, memory, and
individual feedback; and recent agentic recommenders employ collaborative
memory or uncertainty-guided tool delegation.
The remaining distinction is therefore not whether a system uses tools,
memory, feedback, peers, or multiple agents.
It is whether the system treats the evidence boundary, per-source disclosure,
provenance-bearing aggregation, disagreement, and subsequent policy updates
as a coupled, user-governable recommendation process.
PAMR names this coupled process as the research object.
Existing systems may be extended toward PAMR, but no single capability,
external retrieval, user memory, explanation, or multi-agent reasoning, is
sufficient by itself.

Tool-using and travel-planning agents demonstrate the infrastructure needed
for distributed evidence access \citep{openai2025operator,
chen2024travelagent}, and platform shopping assistants illustrate how agentic
interfaces may enter recommendation settings \citep{amazon2024rufus,
amazon2026alexaShopping}.
These systems remain partial PAMR unless source selection, context
disclosure, provenance, and mediation-policy adaptation are explicit,
traceable, and user-governable.

\section{Core Decisions in Personal Agent-Mediated Recommendation}
\label{sec:workflow}

Given the definition in Section~\ref{sec:definition}, PAMR can be viewed as a
constrained information-acquisition problem.
We use the following notation only to clarify the decision space, not to
prescribe a specific algorithm.
Let $u$ have personal agent $\mathcal{A}_u$, history $\mathcal{H}_u$, and
user-specific mediation state $\Theta_u$, including source preferences, trust
estimates, and disclosure policies.
Let $\mathcal{G}$ be a heterogeneous relation graph over agents, services,
platforms, communities, and marketplaces.
Given request $q$, a mediation policy $\pi$ determines the path
\emph{discover} $\rightarrow$ \emph{select} $\rightarrow$ \emph{disclose}
$\rightarrow$ \emph{aggregate} $\rightarrow$ \emph{update}.
At a high level, such a policy can be evaluated by an objective of the form
\begin{align}
\max_{\pi}\quad &
\mathrm{E}\!\left[U(\hat{Y}_u, u, q)\right]
- \lambda_c C(\pi)
- \lambda_p P(\pi) \notag\\
& - \lambda_l L(\pi)
- \lambda_r R(\pi),
\label{eq:mediation_objective}
\end{align}
where $U$ denotes recommendation utility, $C$ query or monetary cost, $P$
privacy and context-disclosure cost, $L$ latency, and $R$ trust, provenance,
or incentive risk.
The point of Eq.~\ref{eq:mediation_objective} is to make explicit what
platform-centric ranking typically leaves implicit: evidence acquisition and
disclosure are recommendation decisions.

\subsection{Source Discovery and Routing}

The evidence-boundary shift makes source discovery part of recommendation.
A personal agent may consider peers, service agents, communities, platforms,
marketplaces, local evidence stores, or referral paths.
The design choice is not simply to ask more sources, but to select sources
under expertise, trust, cost, latency, privacy, incentive, diversity, and
user-control constraints.
This turns source routing into a recommendation variable rather than a
backend retrieval detail.

\subsection{Privacy-Budgeted Context Disclosure}

The decision-locus shift makes context disclosure a recommendation decision.
For each selected source, the agent may send only the user's intent, add task
constraints such as time or budget, or disclose relationship-sensitive
context.
Richer context can improve relevance, but it also increases privacy exposure;
therefore context sharing should be treated as privacy-budgeted translation
rather than prompt engineering.

\subsection{Provenance-Preserving Evidence Aggregation}

The traceability shift makes participant responses evidence channels rather
than final answers.
Responses may contain candidates, claims, rationales, confidence cues, or
referrals.
Aggregation should preserve provenance, disagreement, uncertainty, and
freshness instead of collapsing heterogeneous evidence into an unsupported
ranked list.
Stale, duplicated, or self-promotional evidence may need to be discounted or
surfaced as uncertainty.

\subsection{Feedback-Driven Mediation Adaptation}

The adaptation shift turns presentation and feedback into part of the
mediation loop.
The interface should show enough provenance for the user to inspect which
sources were consulted, what evidence supported or conflicted, and what
context was disclosed.
Feedback should update not only item preferences, but also source preferences,
trust estimates, routing weights, referral paths, and disclosure policies.
This connects PAMR to HCI and generative user interfaces
\citep{amershi2019guidelines, cao2025generativeui}, but only as a mediation
interface rather than a separate agenda.

\section{Source Selection and Disclosure Control}
\label{sec:evaluation}

We use a proof-of-concept study to test whether two mediation
decisions---which sources to query and what context to disclose---affect
recommendation outcomes under a shared ranker.
The study is a partial PAMR instantiation intended to illustrate these
trade-offs, rather than a benchmark or complete system evaluation.

\paragraph{Setup.}
We evaluate 200 hard Yelp restaurant recommendation tasks, each with ten
candidate restaurants, one designated target, and highly rated hard negatives
that violate request constraints.
Two ranking references calibrate task difficulty: Platform Ranking uses only
platform metadata, while Platform LLM Reranker uses the same platform evidence
with an LLM ranking layer.
The mediation conditions form a $2\times2$ comparison that varies evidence
scope---all five sources or a selected top-three subset---and disclosure
policy---full or controlled.
Selective Mediation is a partial PAMR instantiation combining source selection
and controlled disclosure.
All five LLM-based conditions use the same GPT-5.4 ranking pipeline with a
deterministic fallback, so they differ primarily in evidence scope and
disclosure policy.
We report HR@3 for recommendation utility, Trace Validity for whether the
mediation trace satisfies the predefined verification protocol, Exposure for
weighted disclosure burden, and Evidence Calls for source--candidate
evidence requests.

\begin{center}
\scriptsize
\setlength{\tabcolsep}{2.2pt}
\renewcommand{\arraystretch}{1.08}
\resizebox{\columnwidth}{!}{%
\begin{tabular}{lllcccc}
\toprule
\textbf{Condition} & \textbf{Source Scope} & \textbf{Disclosure} &
\textbf{HR@3$\uparrow$} & \makecell{\textbf{Trace}\\\textbf{Validity$\uparrow$}} &
\textbf{Exposure$\downarrow$} & \makecell{\textbf{Evidence}\\\textbf{Calls$\downarrow$}} \\
\midrule
\multicolumn{7}{l}{\textit{Ranking references}} \\
Platform Ranking & Platform & -- & .230 & -- & -- & 0 \\
Platform LLM Reranker & Platform & -- & .675 & -- & -- & 10 \\
\midrule
\multicolumn{7}{l}{\textit{Mediation conditions}} \\
Query-All & All & Full & .690 & .000 & 68.2 & 50 \\
Selection Only & Selected & Full & .715 & .000 & 37.7 & 30 \\
Disclosure Only & All & Controlled & .705 & \textbf{.975} & 14.9 & 50 \\
Selective Mediation & Selected & Controlled & \textbf{.740} & .970 & \textbf{8.8} & \textbf{30} \\
\bottomrule
\end{tabular}
}
\captionsetup{hypcap=false}
\captionof{table}{Diagnostic comparison of source selection and disclosure control. The
five LLM-based conditions use the same GPT-5.4 ranking pipeline and differ
primarily in evidence scope and disclosure policy. Exposure is a within-study
weighted disclosure score; Evidence Calls count source--candidate evidence
requests. Mediation metrics are reported only for multi-source conditions.}
\label{tab:preliminary_pamr}
\end{center}

\paragraph{Findings.}
\textbf{Platform evidence remains limiting.}
The platform-only LLM reranker raises HR@3 from .230 to .675, but selected
and controlled mediation reaches .740.
This suggests an additional benefit from changing what evidence enters the
ranking process.

\textbf{More sources are not always better.}
Source selection improves HR@3 under both controlled disclosure (.705 to .740)
and full disclosure (.690 to .715), while reducing evidence calls by 40\%.
Selective access therefore changes not only cost, but also the observed
recommendation outcome.

\textbf{Disclosure control improves the operating point.}
Controlled disclosure reduces exposure by 77--78\% and produces near-complete
trace validity, without an observed utility penalty.
Selected and controlled mediation thus offers the strongest observed balance
among the multi-source conditions.

\section{Scope Conditions and Failure Modes}
\label{sec:analysis}

Personal agent mediation is most useful when no single platform owns the
relevant evidence boundary.
This includes long-tail, context-sensitive, trust-sensitive, and cross-platform
decisions, such as local services, accessibility-sensitive choices, agent or
service selection, and evidence from communities outside a platform's own
inventory.
Its value is not that more agents are inherently better, but that a user-side
mediator can decide which sources to consult and what context each should
receive, making the evidence boundary itself a controllable part of
recommendation.

\paragraph{When mediation is unnecessary.}
PAMR should not be treated as the default solution for every recommendation
request.
When a single source is authoritative, current, and sufficiently complete,
distributed evidence acquisition may add latency, cost, and disclosure risk
without improving the decision.
The same applies to low-stakes or time-critical requests for which users
prefer a fast local answer over a broader mediation process.
A personal agent should therefore estimate whether the expected value of
additional evidence justifies the added exposure and interaction burden.
It may answer from local or platform evidence, ask the user before expanding
the evidence boundary, or escalate only when sources disagree or confidence
is low.
This selective use is central to PAMR: moving beyond platform-centric ranking
does not mean rejecting platforms, but making reliance on them an explicit
and revisable mediation choice.

The same openness creates risks.
Failures may arise when the mediator misses relevant sources, sends vague
queries, over-discloses context, accepts unsupported or self-promotional
evidence, collapses disagreement into a confident ranking, or reuses stale
local evidence.
Strategic evidence is not hypothetical.
Recent work shows that small, plausible modifications to item descriptions
can manipulate LLM-based rankers \citep{kang2026reliable}.
In a PAMR ecosystem, the attack surface expands because peers, businesses,
marketplaces, and referral agents may all shape the evidence presented to the
mediator.
Provenance must therefore identify not only where a claim originated, but
also who may benefit from it, whether sources are independent, and how
repeated or coordinated evidence should be discounted.
These are central rather than peripheral: every additional evidence path
creates both potential value and new cost, privacy, and incentive risks.
Evaluation should therefore identify which mediation step produced the benefit
or failure; Appendix~\ref{app:failure_taxonomy} provides a fuller taxonomy.

\section{Open Research Challenges}
\label{sec:agenda}

PAMR reframes recommendation as user-side evidence mediation rather than
platform-side item ranking.
Five challenges are especially important.
First, PAMR needs adaptive source routing and provenance-aware aggregation.
Mediators must learn which sources are useful for which requests, how much
context each source should receive, and how conflicting or uncertain evidence
should affect the final recommendation without erasing provenance.

Second, PAMR requires policy learning under user governance.
Future systems must let users inspect and revise disclosure, trust, routing,
and memory policies while still learning from feedback at a practical level of
granularity.
The research problem is how to adapt mediation without turning policy control
into an opaque optimization layer.

Third, PAMR must be robust to strategic sources and evaluated longitudinally.
Sources may manipulate referrals, confidence, or returned evidence, so
robustness cannot be left only to a final reranker.
PAMR makes the manipulation surface larger, because multiple sources may
provide, frame, or withhold evidence, but it also creates new defenses:
source-specific provenance, disclosure logs, disagreement tracking, and
user-correctable mediation policies.
Long-term user studies should test whether people can understand, correct,
and trust these mediation processes over repeated decisions.

Fourth, PAMR requires institutional as well as technical user governance.
A user-facing agent is not necessarily user-controlled: it may still be
hosted, updated, or economically influenced by a platform, marketplace,
operating-system provider, or model vendor.
Future work should therefore study mediation-policy portability, auditable
source access, revocation, interoperable provenance, and separation between
the mediator and evidence providers.
Without such mechanisms, a system may adopt the interface of a personal agent
while preserving platform control over source access and optimization
objectives.
PAMR should be evaluated not only by whether users can edit preferences in an
interface, but also by whether they can move, inspect, and revoke mediation
policies across providers.

Fifth, PAMR needs benchmarks and user studies that evaluate mediation control
itself.
Existing recommender evaluation typically fixes the candidate and evidence
environment, while agent benchmarks emphasize planning, tool use, memory, or
final task success.
PAMR evaluation should instead vary the evidence topology, source
reliability, incentive conflicts, disclosure budgets, and the availability of
alternative providers.
It should distinguish recommendation utility from process properties such as
source coverage, disclosure necessity, provenance completeness, disagreement
preservation, and the user's ability to correct or revoke policy updates.
Counterfactual tests are especially important: would the recommendation
change if a source were removed, if sensitive context were withheld, or if
the user revised a trust rule?
Longitudinal studies should further examine whether users understand these
controls and whether mediation policies remain calibrated as sources,
preferences, and incentives change.
Without such protocols, systems may appear user-centered while the underlying
allocation of evidence control remains unevaluated.

\section{Conclusion}
\label{sec:conclusion}

We have argued that recommender systems should be studied not only as
mechanisms for ranking items, but also as mechanisms for governing the
evidence that produces recommendations.
Personal Agent-Mediated Recommendation relocates this mediation process toward
a user-facing agent that can select evidence sources, control context
disclosure, preserve provenance, and adapt future mediation.
This shift is not equivalent to adding an LLM interface or querying more
tools: it changes who controls the evidence boundary and what recommendation
systems should optimize and evaluate.
Our proof-of-concept results suggest that selective mediation need not trade
away recommendation utility: controlling which evidence enters the ranker and
what context reaches each source yields the strongest observed operating point
among the multi-source conditions.
This remains a partial instantiation rather than a complete PAMR system.
The central claim is therefore not that PAMR is a single finished architecture,
but that user-side evidence mediation should become a first-class research
object in recommender systems.

\section*{Limitations}

This paper is a position paper, not a finished system, deployed agent, or
benchmark.
Its definition and evaluation protocol identify a research space for Personal
Agent-Mediated Recommendation, but they do not prove that any implementation
improves real user outcomes or shifts control in practice.

The proof-of-concept study is intentionally diagnostic and limited.
It uses simulated evidence sources in a single Yelp restaurant domain, with
200 hard tasks, fixed candidate sets, and hand-designed evidence-call budgets.
The evidence sources, disclosure policies, and source-selection rules are
controlled by the experimental setup rather than learned from real users or
deployed services.
The privacy exposure weights are manually specified, and the raw weighted
exposure score should therefore be interpreted as a within-study diagnostic
rather than a dataset-independent privacy measurement.

The study also does not evaluate longitudinal mediation.
Selective Mediation instantiates source selection and disclosure control as a
partial PAMR condition, but does not learn future routing policies from user
feedback, maintain long-term evidence memory, or test whether users can
inspect, correct, or override mediation policies over time.
Results are reported as point estimates; small utility differences should not
be interpreted as statistically reliable without paired uncertainty tests and
cross-dataset validation.

Finally, the current evidence does not establish that user-facing agents
necessarily become user-owned or user-governable.
Real deployments may still be controlled by platforms, enterprises, or
marketplaces.
Our claim is therefore conceptual and diagnostic: PAMR identifies the control
problem that recommender systems should study, but this paper does not yet
demonstrate that deployed agents solve it.

\bibliography{custom}

\appendix
\section{Trace Components}
\label{app:benchmark_schema}

Useful PAMR traces may include the user request, discovered candidate
participants, selected participants, plausible but unselected participants,
context sent to each participant, privacy exposure estimates, participant
responses, provenance, referrals, aggregation rationale, user-facing
presentation, feedback signals, and optional profile, trust, evidence-memory,
or routing updates.
We leave concrete schema design to future benchmark work because the right
fields will depend on the application domain and deployment setting.

\section{Failure Taxonomy}
\label{app:failure_taxonomy}

PAMR failures can be grouped by where they enter the mediation loop.
\textit{Discovery and routing failures} include missing relevant participants,
retrieving irrelevant ones, relying on a narrow or hub-dominated relation
graph, or misallocating budget.
\textit{Context failures} include vague participant queries or excessive
sharing of user information.
\textit{Response failures} include hallucinated candidates, unsupported
rationales, duplicate suggestions, noisy referrals, stale participant
profiles, self-promotion, and manipulative referrals.
\textit{Aggregation failures} include constraint violations, loss of
provenance, stale local evidence, and over-weighting trusted but inaccurate
sources.
\textit{Presentation and adaptation failures} include over-persuasive
interfaces, hidden uncertainty, erroneous feedback extraction, and over-updated
trust or routing policies.
Referral can expand coverage when the known participant set is insufficient,
but it should be depth-limited and trust-aware to avoid runaway cost, noisy
paths, manipulation, or feedback loops that invite low-quality agents into the
mediation path.
This taxonomy is intended as a diagnostic checklist rather than a closed set
of error categories.

\section{Additional Metrics}
\label{app:additional_metrics}

Beyond the minimal protocol in Section~\ref{sec:evaluation}, PAMR evaluations
may track additional dimensions such as outcome quality, discovery and routing
quality, cold-start efficiency, temporal stability, and incentive
compatibility.
Concrete measures may include constraint satisfaction, task success, useful
participant recall, irrelevant query rate, referral usefulness,
interactions-to-useful-routing, participant calibration speed, routing
stability, trust calibration, stale evidence rate, self-promotion rate,
manipulative referral rate, and robustness to strategic participants.
These metrics should be treated as diagnostic lenses rather than a fixed
scoring scheme.

\section{Proof-of-Concept Study Setup}
\label{app:study_setup}

The diagnostic study uses a 200-task hard subset of AgentRecBench-Yelp.
Each task contains a structured restaurant request, ten candidate restaurants,
a single gold target satisfying all constraints, mediation-profile metadata,
and hard negatives that are highly rated but violate at least one request
constraint.
The gold target is intentionally non-obvious: under pure platform rating, its
average rank is 5.49 out of 10, and it appears in the top three only 23\% of
the time.

The source environment contains five simulated evidence sources: platform
metadata, recent reviews, community reviews, peer-like signals, and
business-owned information.
Selective Mediation chooses roughly three sources per task using relevance,
trust, and cost, and then applies persona-specific disclosure rules.
This setup instantiates source selection and disclosure control, but it does
not evaluate longitudinal feedback-driven adaptation.
LLM-based ranking and explanation conditions used GPT-5.4 through the OpenAI
API \citep{openai2026gpt54}.

\section{Baseline Conditions}
\label{app:baseline_conditions}

\noindent\textbf{Platform Ranking.}
Ranks candidates by platform metadata, primarily rating, and performs no
external evidence mediation.

\noindent\textbf{Platform LLM Reranker.}
Uses platform evidence with constraint-aware LLM reranking and explanation.
It calibrates ranking utility but does not perform source selection or
context disclosure.

\noindent\textbf{Query-All.}
Queries all five sources for every candidate with full context disclosure.

\noindent\textbf{Selection Only.}
Selects a subset of sources but uses full context disclosure.

\noindent\textbf{Disclosure Only.}
Queries all five sources but applies persona-specific disclosure rules.

\noindent\textbf{Selective Mediation.}
Selects a subset of sources and applies persona-specific disclosure rules.
This partial PAMR instantiation combines the two mediation decisions studied
in Section~\ref{sec:evaluation}.

\section{Metric Definitions}
\label{app:metric_definitions}

\noindent\textbf{HR@3.}
HR@3 is 1 if the gold target appears in the top three recommendations for a
task and 0 otherwise; reported values are task averages.

\noindent\textbf{Trace Validity.}
Trace Validity corresponds to Mediation Trace Validity (MTV).
It is computed per task as a binary verifier:
\[
\mathrm{MTV}(t)=\mathbf{1}[\mathrm{CV}\wedge \mathrm{ES}\wedge \mathrm{DC}\wedge \mathrm{TC}],
\]
where CV checks candidate validity, ES checks evidence support for top-ranked
items, DC checks disclosure compliance with the persona policy, and TC checks
trace completeness.
Reported MTV is the average over tasks.
More concretely, CV requires all ranked items to belong to the candidate set;
ES requires top-ranked items to have supporting evidence rather than only
contradictory claims; DC requires disclosed slots to obey the persona policy;
and TC requires the ranking, explanation, source-call records, evidence, and
disclosure log to be present when mediation is performed.
Values near 0 or 1 reflect task averages under a fixed protocol, not a
method-level flag.

\noindent\textbf{Exposure.}
Exposure is a within-study weighted disclosure score.
It sums exposure over all disclosed user slots:
\[
\mathrm{PES}=\sum_{(s,v)\in D} \mathrm{sensitivity}(s)\cdot \mathrm{risk}(v),
\]
where $s$ is a disclosed slot, $v$ is the receiving source, and $D$ is the
set of disclosures.
Slot sensitivity weights are low=1, medium=3, and high=5.
Source-risk weights are 1.0 for platform and recent-review sources, 1.2 for
community-review sources, and 1.5 for peer-like and business-owned sources.
The reported values should be interpreted as within-study diagnostics rather
than dataset-independent privacy measurements.

\noindent\textbf{Evidence Calls.}
An Evidence Call is one source--candidate evidence request to a platform,
external, or simulated evidence source.
In this setup, each task has ten candidate restaurants and five multi-source
evidence channels, so query-all conditions issue $5\times10=50$ such requests
per task, while source-selection conditions issue $3\times10=30$.
Platform LLM Reranker uses one platform evidence request per candidate, for
ten Evidence Calls per task.
LLM inference calls are not counted in this metric.

\noindent\textbf{Additional diagnostic metrics.}
Several diagnostics are useful for implementation analysis but are omitted
from the main text to keep the position claim focused.
CS@1 measures whether the top-ranked recommendation satisfies all structured
request constraints.
Average disclosed slots counts the number of user-context fields sent to
sources per task, without weighting by sensitivity or source risk.
Unsupported claim rate measures the share of evidence claims not supported by
the verifier.
Disclosure violation rate measures the fraction of tasks in which the
disclosure policy is violated.
Minimum Necessary Disclosure Rate (MNDR) is the fraction of disclosed slots
that are necessary for the request and source.
Sensitive Disclosure Violation Rate (SDVR) measures tasks containing
unnecessary medium- or high-sensitivity disclosure.
Privacy-Aware Utility (PAU) combines utility and normalized exposure:
\[
\mathrm{PAU}(\lambda)=\mathrm{HR@3}-\lambda\cdot \mathrm{normalized\ exposure}.
\]
These diagnostics help debug mediation behavior, but they are not required
for the central comparison in Table~\ref{tab:preliminary_pamr}.

\section{Full Results and Implementation Mapping}
\label{app:full_results}

Table~\ref{tab:full_results} reports the full diagnostic result table for the
proof-of-concept study.
The study reports point estimates; small utility differences should not be
interpreted as statistically reliable without paired uncertainty tests such as
bootstrap confidence intervals, permutation tests, or McNemar tests.
The purpose of the table is to expose how source scope and disclosure policy
change the observed utility--validity--privacy--cost operating point.

\begin{table*}[t]
\centering
\scriptsize
\setlength{\tabcolsep}{4pt}
\renewcommand{\arraystretch}{1.08}
\begin{tabular}{lllcccc}
\toprule
\textbf{Condition} & \textbf{Source Scope} & \textbf{Disclosure} &
\textbf{HR@3} & \makecell{\textbf{Trace}\\\textbf{Validity}} &
\textbf{Exposure} & \makecell{\textbf{Evidence}\\\textbf{Calls}} \\
\midrule
Platform Ranking & Platform & -- & .230 & -- & -- & 0 \\
Platform LLM Reranker & Platform & -- & .675 & -- & -- & 10 \\
Query-All & All & Full & .690 & .000 & 68.2 & 50 \\
Selection Only & Selected & Full & .715 & .000 & 37.7 & 30 \\
Disclosure Only & All & Controlled & .705 & .975 & 14.9 & 50 \\
Selective Mediation & Selected & Controlled & .740 & .970 & 8.8 & 30 \\
\bottomrule
\end{tabular}
\caption{Full diagnostic results. Exposure is a within-study weighted
disclosure score. Evidence Calls count source--candidate evidence requests
and exclude LLM inference calls. Mediation metrics are reported only for
multi-source conditions.}
\label{tab:full_results}
\end{table*}

\noindent\textbf{Source selection.}
Comparing Query-All with Selection Only shows the effect of reducing the
evidence scope under full disclosure: HR@3 rises from .690 to .715, Exposure
drops from 68.2 to 37.7, and Evidence Calls drop from 50 to 30.
Comparing Disclosure Only with Selective Mediation shows the same
source-selection contrast under controlled disclosure: HR@3 rises from .705
to .740, Exposure drops from 14.9 to 8.8, and Evidence Calls drop from 50 to
30.

\noindent\textbf{Disclosure control.}
Comparing Query-All with Disclosure Only shows the effect of disclosure
control under all-source querying: Trace Validity rises from .000 to .975 and
Exposure drops from 68.2 to 14.9.
Comparing Selection Only with Selective Mediation shows the same
disclosure-control contrast under selected-source querying: Trace Validity
rises from .000 to .970 and Exposure drops from 37.7 to 8.8.
In both source-scope settings, controlled disclosure satisfies the predefined
trace-validity protocol without an observed utility penalty.

\begin{table}[t]
\centering
\scriptsize
\setlength{\tabcolsep}{3pt}
\renewcommand{\arraystretch}{1.08}
\resizebox{\columnwidth}{!}{%
\begin{tabular}{ll}
\toprule
\textbf{Paper condition} & \textbf{Implementation condition} \\
\midrule
Platform Ranking & \texttt{platform\_only} \\
Platform LLM Reranker & \texttt{instruct\_reranker} \\
Query-All & \texttt{query\_all\_full} \\
Selection Only & \texttt{selective\_no\_disclosure} \\
Disclosure Only & \texttt{selective\_no\_selection} \\
Selective Mediation & \texttt{selective\_pamr} \\
\bottomrule
\end{tabular}
}
\caption{Mapping between paper-facing condition names and implementation
identifiers used in trace and metric files.}
\label{tab:condition_mapping}
\end{table}

\end{document}